\documentclass[fleqn,usenatbib]{mnras}

\usepackage{newtxtext,newtxmath}

\usepackage[T1]{fontenc}

\DeclareRobustCommand{\VAN}[3]{#2}
\let\VANthebibliography\thebibliography
\def\thebibliography{\DeclareRobustCommand{\VAN}[3]{##3}\VANthebibliography}

\usepackage{graphicx}	
\usepackage{amsmath}	

\newcommand{\src}{J1755$-$2527}
\newcommand{\srcfull}{ASKAP J175534.9$-$252749.1}

\DeclareMathOperator{\erfcx}{erfcx}
\DeclareMathOperator{\erfc}{erfc}
\DeclareMathOperator{\erf}{erf}
\DeclareMathOperator{\emg}{emg}

\newcommand{\deriv}[2]{\frac{{\rm d}{#1}}{{\rm d}{#2}}}
\newcommand{\dd}[2]{\frac{{\rm d^2}{#1}}{{\rm d}{#2}^2}}
\newcommand{\ToA}[1]{{\rm ToA}_{\rm {#1}}}

\newcommand{\Fig}{Fig.}

\newcommand{\Figs}{Figs.}

\newcommand{\Sect}{Section}

\newcommand{\Tab}{Table}

\newcommand{\eqn}{equation}

\newcommand{\Eqn}{Equation}

\defcitealias{2024MNRAS.535..909D}{Paper~I}

\title[1.16-hr-repeating pulses from \src{}]{A new long period radio transient: Discovery of pulses repeating every 1.16 hours from \srcfull{}}

\author[S. J. McSweeney et al.]{
Samuel J. McSweeney$^{1}$,
Natasha Hurley-Walker$^{1}$,
Csan\'{a}d Horv\'{a}th$^{1}$,
Akash Anumarlapudi$^{2}$,
\newauthor
Angie Waszewski$^{1,5}$,
Dougal Dobie$^{3,4}$,
David L.~Kaplan$^{2}$,
John Morgan$^{5}$,
Kovi Rose$^{3,6}$,
and Ziteng Wang$^{1}$
\\
$^{1}$International Centre for Radio Astronomy Research, Curtin University, Bentley, WA 6102, Australia\\
$^{2}$Department of Physics, University of Wisconsin-Milwaukee, P.O. Box 413, Milwaukee, WI 53201, USA\\
$^{3}$Sydney Institute for Astronomy, School of Physics, The University of Sydney, NSW 2006, Australia\\
$^{4}$ARC Centre of Excellence for Gravitational Wave Discovery (OzGrav), Hawthorn, VIC 3122, Australia\\
$^{5}$CSIRO Space and Astronomy, P.O. Box 1130, Bentley, WA 6102, Australia\\
$^{6}$Australian Telescope National Facility CSIRO, P.O. Box 76, Epping, NSW 1710, Australia\\
}

\date{Accepted XXX. Received YYY; in original form ZZZ}

\pubyear{\the\year{}}

\begin{document}
\label{firstpage}
\pagerange{\pageref{firstpage}--\pageref{lastpage}}
\maketitle

\begin{abstract}
We report the discovery of several new pulses from the source \srcfull{} (\src{}), originally identified from a single 2-min long pulse, confirming it as a long period transient (LPT) with a period of ${\sim}1.16\,$hours. The pulses are significantly scattered, consistent with Galactic electron density models. Two of the new pulses also had measurable polarisation, but unlike the originally detected pulse, the polarisation angle does not behave as expected from the rotating vector model. We interpret historical non-detections of \src{} as an intrinsic intermittency that occurs on month-long timescales, and discuss possible causes. We conjecture that, like some other LPTs with periods $\gtrsim 1\,$hour, \src{} may host a white dwarf in a binary orbit, but note that its period is marginally shorter than the canonical orbital period minimum of cataclysmic variables. Our work highlights the importance of additional observations in establishing the nature of unusual radio-emitting objects.
\end{abstract}




\section{Introduction} \label{sec:introduction}

\srcfull{} (hereafter, \src{}) is a transient radio source \citep[][hereafter \citetalias{2024MNRAS.535..909D}]{2024MNRAS.535..909D} discovered in the Variables and Slow Transients (VAST) survey \citep{2013PASA...30....6M}.
A single, highly polarised pulse was observed, lasting approximately 2 minutes, exhibiting a steep spectrum ($\alpha = -3.1$, for flux density $S_\nu\propto\nu^\alpha$).
Extensive follow up in new and archival data at multiple wavelengths turned up no other pulses.
Despite this, the authors concluded that \src{} was most likely a long period transient (LPT), a newly emerging class of radio source with periods ranging from minutes to hours \citep[see e.g.][]{2022Natur.601..526H,2023Natur.619..487H,2024NatAs...8.1159C}.
Without additional evidence, however, \src{} has heretofore not been classified as an LPT in current catalogs.

Although optical observations have provided compelling evidence that at least some LPTs are detached, possibly strongly magnetic, white dwarf / M-dwarf (WD + MD) binary systems \citep{deRuiter2025,2024ApJ...976L..21H,2025A&A...695L...8R}, it remains unclear if this model can account for all known LPTs \citep[e.g.][]{2022ApJ...940...72R,Lee2025}.
Optical follow-up is challenging for \src{}, which is situated very close to the Galactic plane ($b = -0\overset{\circ}{.}12$), at a distance of ${\sim}4.7\,$kpc \citepalias{2024MNRAS.535..909D}.
Unsurprisingly, archival searches at \src{}'s location did not turn up any source down to ${\sim}23$ AB mag.
Without more detections in the radio or at other wavelengths, it seemed unlikely that the nature of \src{} could ever be resolved.

In this paper, we present new radio observations (spanning frequencies from $154\,$MHz to $3.1\,$GHz) in which \src{} was re-detected, confirming that it is indeed an LPT with a period of ${\sim}1.16\,$hours.
In \S\ref{sec:observations}, we describe the observations themselves, some of which were taken as part of independent surveys, and some of which were taken as dedicated follow-up of \src{} once it became apparent that it was still active.
In \S\ref{sec:analysis} we present a timing analysis of the new detections, as well as revised estimates of the dispersion measure (DM) and the scattering timescale.
We also include new polarisation detections in which the polarisation angle (PA) behaviour differs significantly from the originally detected pulse.
In \S\ref{sec:discussion} we revisit the question of the nature of \src{}, arguing in favour of a white dwarf interpretation.
The main conclusions are summarised in \S\ref{sec:summary}.

\section{Observations} \label{sec:observations}

Pulses from \src{} were detected with the Murchison Widefield Array \citep[MWA;][]{Tingay2013}, the Australian SKA Pathfinder \citep[ASKAP;][]{2021PASA...38....9H}, MeerKAT \citep{2016mks..confE...1J}, and the Australia Telescope Compact Array \citep[ATCA;][]{cabb}, all radio interferometers capable of imaging.
A summary of all the observations containing pulse detections included in this work is presented in \Tab~\ref{tbl:obs}.
In this section, we describe how each telescope was used to generate dynamic spectra of \src{}.

\begin{table*}
  \centering
  \caption{Observations summary}
  \label{tbl:obs}
  \begin{tabular}{lllcccc}
    \hline
    Date range & Telescope & Project & $\nu$ & $\Delta\nu$ & Obs. (num. $\times$) & Number of \\ 
    (MJD) & & & (MHz) & (MHz) & length (min) & pulses \\
    \hline 
    59965 & ASKAP & SB47253 (original discovery) & 887.5 & 288 & 12.1 & 1 \\
    59966 & MWA & G0060 & 161.92 & 15.36 & 6.7 & 1 \\
    60040  & ASKAP & SB49153 & 887.5 & 288 & 12.1 & 1 \\
    60092 - 60093 & MeerKAT & DDT-20230525-DD-01 & 1284 & 856 & 307 & 4 \\
    60463 - 60563 & MWA & G0080 & 200.32 & 30.72 & $16 \times 4.9$ & 16 \\
    60503 & ASKAP & SB63600 & 887.5 & 288 & 12.3 & 1 \\
    60572 & MWA & D0042 & 184.96 & 30.72 & $4 \times 4.9$ & 4 \\
    60587 & ATCA & C3363 & 2100 & 2048 & 93.2 & 2 \\
    60592 - 60602 & MWA & D0042 & 200.32 & 30.72 & $17 \times 4.9$ & 17 \\
    60602 & MeerKAT & DDT-20241015-NH-01 & 812.8 & 495.3 & 10 & 1 \\
    \hline
  \end{tabular}
\end{table*}

\subsection{MWA} \label{sec:mwa}

The MWA, in its Phase~\textsc{ii} ``extended'' configuration \citep[][]{2018PASA...35...33W}, was used to scan the Galactic plane for transient radio sources under the `Galactic Plane Monitor', project code G0080. These observations were conducted over June--September\,2022, and June\,2024--March\,2025. Relevant to this work, the region $285^\circ < l < 65^\circ$, $|b| < 15^\circ$ was scanned on a bi-weekly cadence at 185--215\,MHz in both June--September\,2022 and 2024 \citep[see Methods of][a full description will be released by Hurley-Walker et al., in prep]{2023Natur.619..487H}. In this region, the root-mean-squared (RMS) noise level on the processed four-second timescale is $\sim$80\,mJy\,beam$^{-1}$. Amongst several detection methods (described in full by Horv\'ath et al., submitted, a convolution filter is employed to extract pulse-like signals. In June--September 2022, no pulses from \src{} were detected, but in June 2024, the source was blindly detected as a ${\sim}70$-s-wide pulse whose peak flux density was 665\,mJy.

Subsequent to its redetection, we commenced a Director's Discretionary Time (DDT) campaign under project code D0042, to obtain further observations, adjusting the frequency from 200.32\,MHz to 184.96\,MHz for some of the observations to attempt to improve measurements of its DM.
Time of arrival (ToA) predictions for the DDT observations were made using a preliminary ephemeris derived from the GPM detections, which, by virtue at being at the same or similar frequencies, absorbed the delays due to dispersion and scattering into the `PEPOCH' term.

We imaged both the 2024 GPM data and the follow-up DDT data at the location of \src{} using \textsc{WSClean} \citep{2014MNRAS.444..606O}, masking \src{} itself, forming a deep model of the sky for each observation. After subtracting this from the visibilities, we phase-rotated to \src{}, and averaged the baseline data to produce a dynamic spectrum.

An additional MWA observation made as part of the MWA Interplanetary Scintillation (IPS) Survey, project code G0060, taken just one day after the original ASKAP pulse, was found to have \src{} in the field of view.
The pulse was detected as a point source in the image formed from the third of three adjacent 200-second, near-Sun observations (30$^{\circ}$ elongation).
In this paper we present both the second and third observations (6.7 minutes total), in order to make the contrast between detection and non-detection more visible in the lightcurve formed by averaging over the 15.36-MHz subband, using 10-second time bins.
A full description of how MWA IPS observations are scheduled and processed can be found in Section~2 of \citet{Morgan2022}, but the essential details are summarised here.
The observations are taken over two equal bands of frequency, centred on approximately 80\,MHz and 160\,MHz.
For this study, only the upper band has been processed. 
The observations were self-calibrated against a sky model based on the GLEAM survey \citep{Hurley-Walker2017} using \textsc{mwa\_hyperdrive}\footnote{\url{https://github.com/MWATelescope/mwa_hyperdrive}}, and then imaged using \textsc{WSClean}.
Further details of the imaging of IPS observations since the first data release can be found in Waszewski et al. (submitted).

\subsection{ASKAP} \label{sec:askap}

\citetalias{2024MNRAS.535..909D} presented an exhaustive search of the ASKAP data available at the time of writing, comprising 60\,h of data, and yielding only a single detection of \src{} (in Scheduling Block ID 47253). Subsequent to the writing of \citetalias{2024MNRAS.535..909D}, a further search for transient signals in ASKAP data was performed (SBID 63600) in which \src{} produced a detectable pulsation.

With the re-detection of the source in the MWA data, and the calculation of an ephemeris (\Sect~\ref{sec:analysis}), we re-reduced the data in which we would predict that \src{} would appear. We made one further detection, in SBID~49153, at a relatively low significance ($\sim5\sigma$).
We followed a similar procedure as the above to generate dynamic spectra for each of the observations, using a minimum baseline length of 500\,m to suppress uncleaned Galactic diffuse emission.

\subsection{MeerKAT} \label{sec:meerkat}

MeerKAT observations were undertaken on 2024-10-19, under proposal code DDT-20241015-NH-01 and capture block 1729341386, using the UHF band spanning 544 -- 1088\,MHz.
As well as correlator observations undertaken at 2-s/132.812-kHz resolution, we also employed the Pulsar Timing User Supplied Equipment \citep[PTUSE;][]{2020PASA...37...28B} in search mode, using 37.45-$\mu$s sampling. The observation included typical bandpass, polarisation, and phase calibrators, as well as phase-up and test pulsar observations. \src{} was tracked from 13:54:36 to 14:04:35 UTC.

The correlator data were calibrated using the standard SARAO SDP calibration pipeline, and imaged using \textsc{WSClean}. We formed a deep model of the sources within the primary beam, excepting \src{}, and subtracted this from the visibilities.
We then averaged the visibilities, excluding all baselines shorter than 150\,m, to produce a dynamic spectrum of the source, shown in \Fig~\ref{fig:stacked_spectra}.
We used the source-finding software \texttt{Aegean} \citep{2018PASA...35...11H} to obtain an updated position of 17:55:34.87(2) $-$25:27:49.9(4).
As in \citetalias{2024MNRAS.535..909D}, we find no coincident sources in either optical or near infrared at the same location (see Appendix~\ref{app:localization}).

We also re-imaged the original 2023-05-28 L-band MeerKAT observation described in \citetalias{2024MNRAS.535..909D}, which was taken under project code DDT-20230525-DD-01 under capture block 1685306788. We performed a similar processing of the data, and identified four pulses with ToAs consistent with an ephemeris derived from the other data. The folded and stacked dynamic spectra are shown in \Fig~\ref{fig:stacked_spectra}.

\subsection{ATCA}

Following its re-detection in the ASKAP data, follow-up observations were conducted at L/S bands with the ATCA telescope under the project code C3363 (P.I: Tara Murphy). \src\ was observed once on 2024-10-04, for a total of 1.5\,hrs, and again on 2024-10-12, for a total of 2.4\,hrs. At the beginning of each observation, 1934$-$638 was used to calibrate the bandpass, and during the observations, 1748$-$253 was used as a phase calibrator. Data were flagged, calibrated, and imaged using standard recipes from the Common Astronomy Software Applications package \citep[CASA;][]{casa}. The resulting sky model was subtracted from the visibilities (excluding the source of interest). These model-subtracted visibilities were phase-centered on the \src{} and averaged over the baselines to produce the dynamic spectra. 

We identified two faint pulses (4--5 $\sigma$; see Table~\ref{tbl:toas}) in the observation taken on 2024-10-04. However, observations taken on 2024-10-12 did not yield any pulses (at 3-$\sigma$ level). We inspected the data around the expected time of arrival (using Equation~\ref{eqn:timing}) but did not find any pulses. The instantaneous noise level, per 10\,s integration time bin, around the expected time of arrivals is close to 1\,mJy\,beam$^{-1}$. If pulse-to-pulse flux density variations, similar to ones observed in MWA data (Table~\ref{tbl:toas}), also occur at L/S band, then even a factor 1.5--2 variation can be sufficient to explain the detection of pulses on 2024-10-04 and the absence of pulses on 2024-10-12, given that the source is faint at higher frequencies. 

\section{Analysis \& results} \label{sec:analysis}

\subsection{Pulse morphology}

All observed pulses consist of a single burst, approximately Gaussian in shape, with the sole exception of the MeerKAT pulse observed on 2024-10-19, for which two bursts were observed separated by ${\sim}125\,$s.
Generally, the low-frequency (MWA) pulses tend to
\begin{itemize}
    \item be wider than their higher frequency counterparts,
    \item exhibit an asymmetry resembling a scatter-broadened tail, and
    \item have minimal pulse-to-pulse morphological variation.
\end{itemize}

The low-frequency asymmetry is consistent with a scattering time scale of a few tens of seconds at MWA frequencies, which scales to a few tens of milliseconds at 1\,GHz.
This is in broad agreement with the model predictions of NE2001 ($\tau_{\rm sc,1GHz} = 66\,$ms) and YMW16 ($\tau_{\rm sc,1 GHz} = 75\,$ms) at the source's location assuming the DM value of 710\,pc\,cm$^{-3}$ reported in \citetalias{2024MNRAS.535..909D}.

\begin{figure*}
      \centering
          \includegraphics[width=0.95\linewidth]{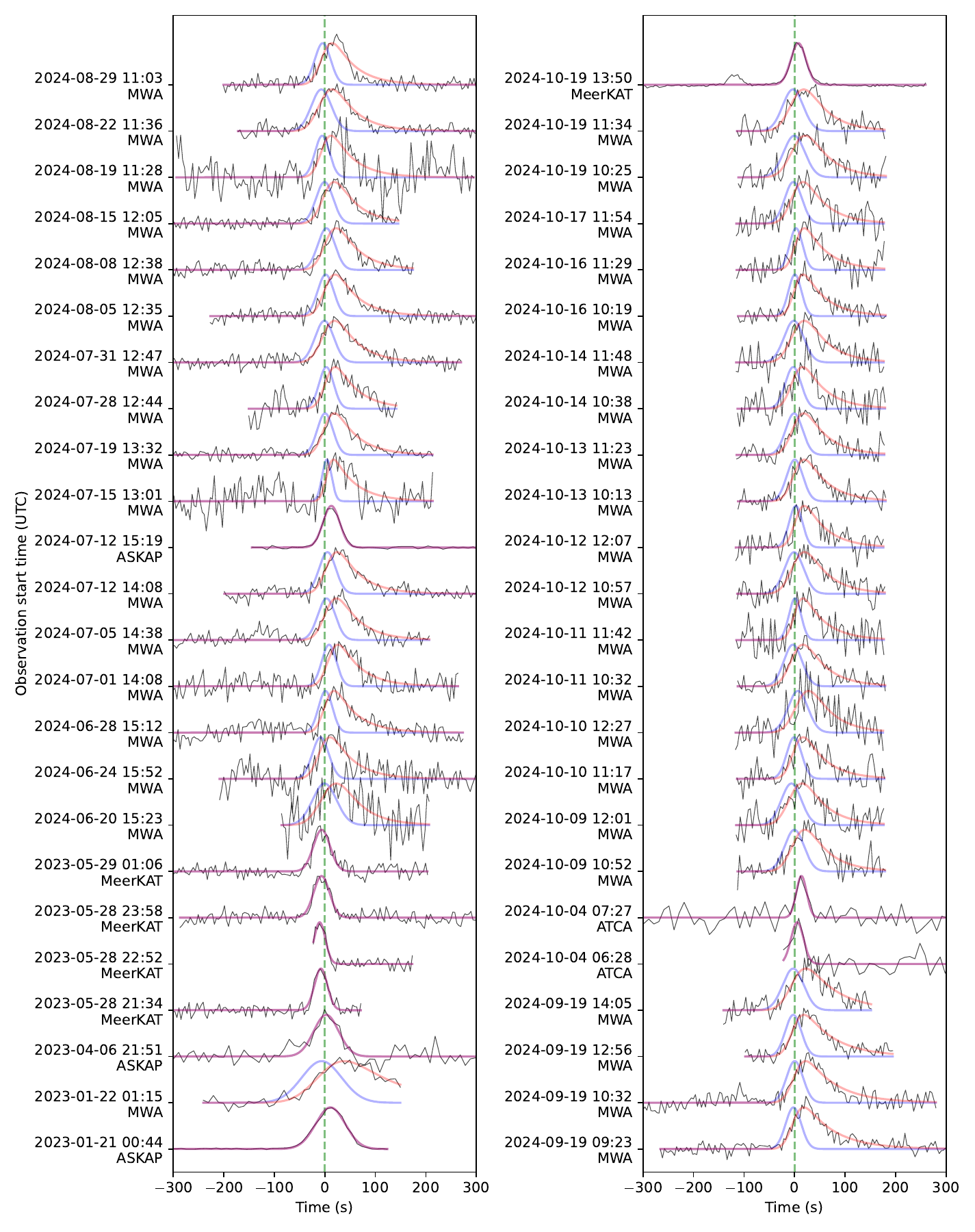}
              \caption{Pulsestack of barycentred, dedispersed profiles from the ephemeris. The dashed green vertical line marks zero phase. Baselines have been fitted and subtracted from each lightcurve. The red curves are fits of the normalised pulses to \Eqn~\ref{eqn:emg}, and the blue curves are the inferred Gaussian pulses without scattering, also normalised (see main text for details).}
                  \label{fig:pulsestack}
\end{figure*}

To account for the asymmetry, we fit exponentially modified Gaussians to each pulse with the form,
\begin{equation}
  \emg(t) = A \frac{\sigma}{\tau}\sqrt{\frac{\pi}{2}}
    \exp\left(-\frac12 \left( \frac{t - \mu}{\sigma} \right)^2 \right)  \erfcx \left(\frac{1}{\sqrt{2}} \left(\frac{\sigma}{\tau} - \frac{t - \mu}{\sigma} \right) \right),
  \label{eqn:emg}
\end{equation}
where $\erfcx(X) = \exp(X^2) \erfc(X)$ is the scaled complementary error function; $A$, $\mu$, and $\sigma$ are the height, location, and scale of an unscattered Gaussian pulse; and the scattering time scale was fixed at
\begin{equation}
      \tau(\nu) = 0.07\, \left(\frac{\nu}{\rm GHz}\right)^{-4} \, {\rm s},
\end{equation}
according to each observation's centre frequency.
A pulse stack containing all observed pulses and the corresponding unscattered pulses derived from the fits is shown in \Fig~\ref{fig:pulsestack}.

For the MeerKAT pulse with two components, we fit only the brighter component.
We checked the higher time resolution PTUSE data for finer substructure, but did not find anything smaller than the components that can be seen in \Fig~\ref{fig:pulsestack}.

\subsection{Timing solution} \label{sec:timing}

Since the large scattering time scale at low frequencies significantly distorts the pulse shapes, common methods for defining ToAs can potentially result in systematic errors of up to tens of seconds, as discussed in detail in Appendix \ref{app:scattering_dm}.
These systematic errors set up a degeneracy between DM and scattering time scales that is not easily disambiguated.

\begin{figure}
      \centering
          \includegraphics[width=0.95\linewidth]{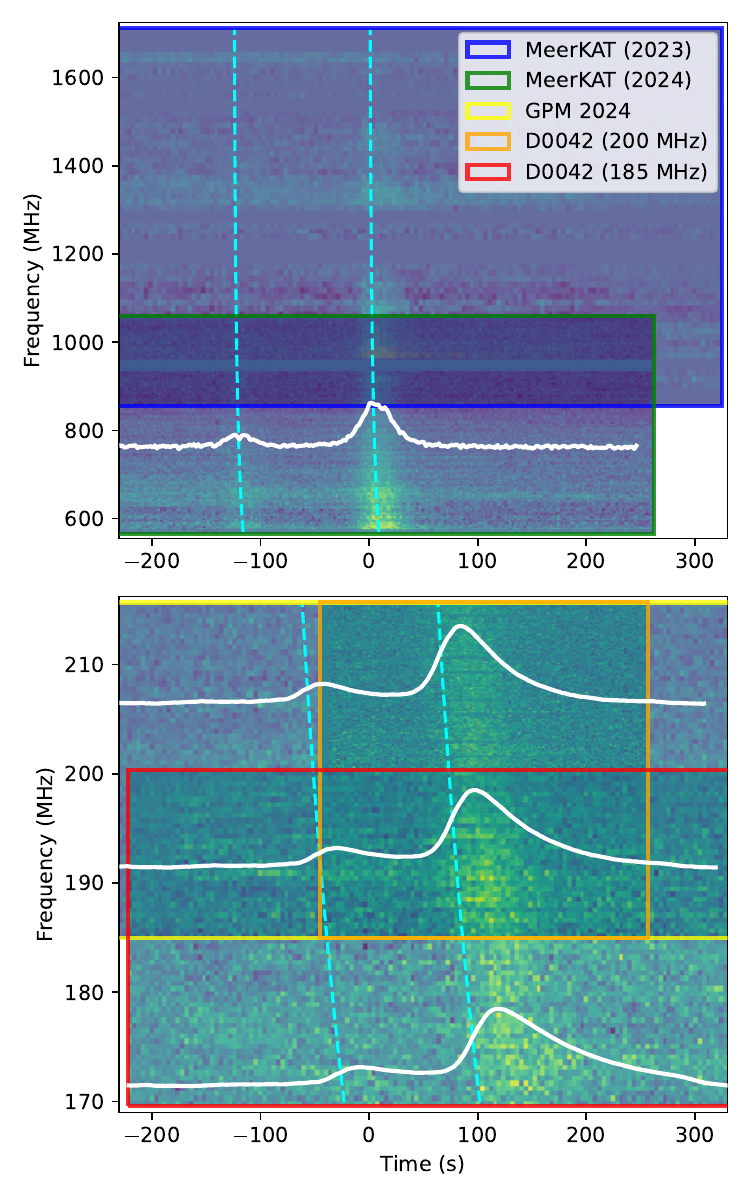}
              \caption{Stacked (semi-transparent) dynamic spectra, where each rectangle indicates the observing campaign whose spectra were barycentred and folded according to the ephemeris. The spectra have not been dedispersed, but the cyan dashed lines indicate where the two components seen in the MeerKAT (2024) observation would appear due to dispersion. The white curve in the top panel is the dedispersed profile of the MeerKAT (2024) observation, and the white curves in the bottom panel are the same profile subjected to scattering with $\tau_{\rm sc} = \tau_{\rm sc,1\,GHz} (\nu/{\rm 1\,GHz})^{-4}$ at frequencies 175, 195, and 210 MHz, where $\tau_{\rm sc,1\,GHz} = 70$\,ms.}
                  \label{fig:stacked_spectra}
\end{figure}

This is demonstrated in \Fig~\ref{fig:stacked_spectra}, which shows stacked MWA and MeerKAT spectra.
The cyan lines show the expected ToAs across the whole observed frequency range if one assumes only a DM measured at higher frequencies, where scattering is negligible.
The scattering predicted by Galactic electron density models, however, is sufficient to account for the apparent extra delay, as is evidenced by the relatively good agreement with the modelled unscattered pulses with the expected ToAs derived from the ephemeris, illustrated in the pulsestack shown in \Fig~\ref{fig:pulsestack}.

We therefore identified the ToAs with the $\mu$ parameter in the fits given in \Eqn~\ref{eqn:emg}, which represents the location of the assumed Gaussian pulse before scattering.
Because the scattering time scale could not be reliably measured for all individual pulses, we obtained ToAs by assuming a fixed scattering time of $\tau_{\rm sc,1 GHz} = 70$\,ms.

\begin{figure*}
  \centering
  \includegraphics[width=0.98\linewidth]{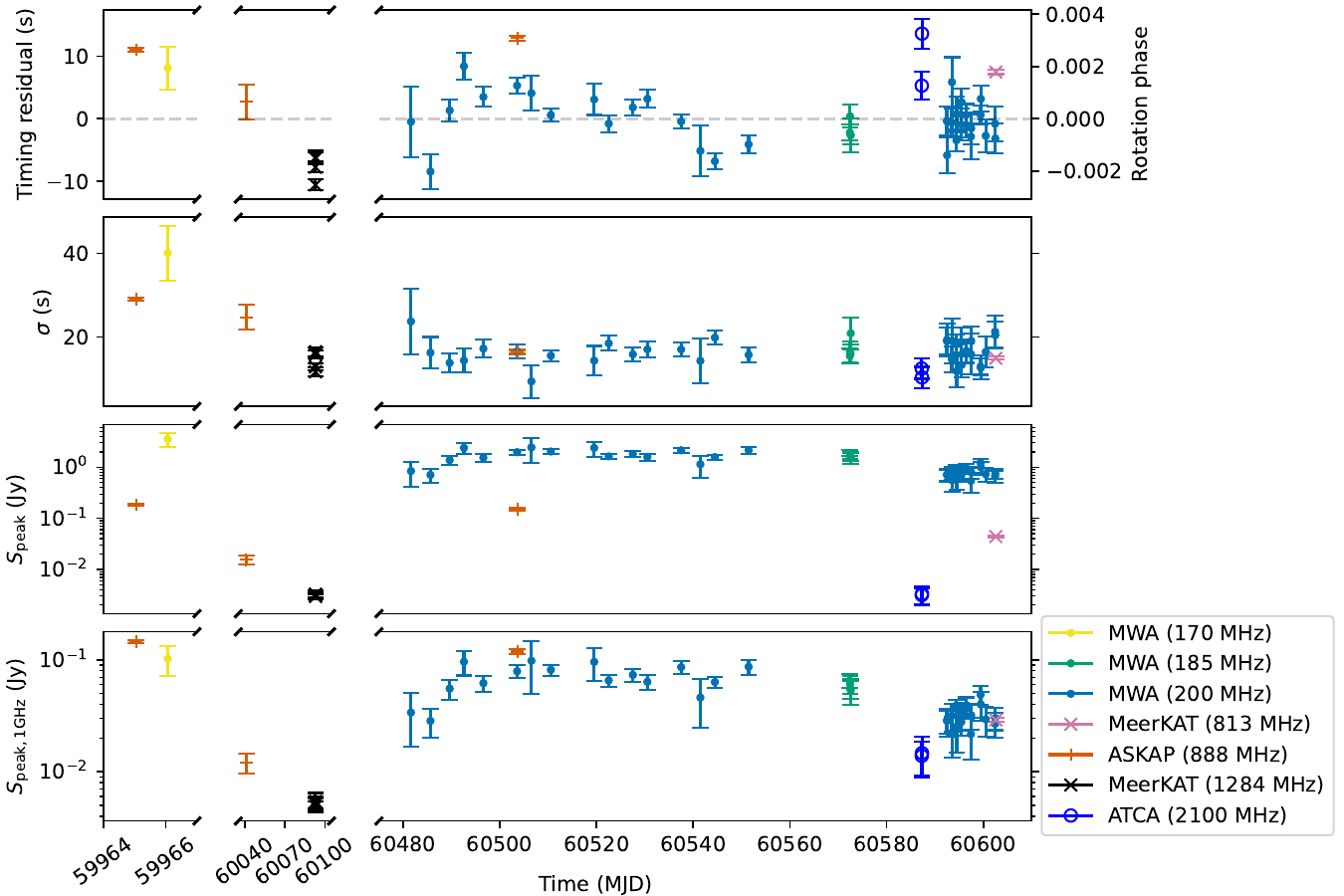}
  \caption{Timing residuals (top panel) and other pulse properties derived from the fits of \Eqn~\ref{eqn:emg} to the individual light curves (second and third panels). The peak flux, $S_{\rm peak}$, equivalent to the $A$ parameter in \eqn~\ref{eqn:emg}, is the peak flux of the modelled unscattered pulse. The bottom panel displays $S_{\rm peak,1 GHz} = S_{\rm peak} (\nu/{\rm GHz})^{-\alpha}$, where $\alpha = -2$ has been assumed. Notable features of interest include the relative brightness of the early 2023 detections (leftmost two points), and the relative stability of the pulse fluences (and pulse morphology) of the MWA detections in the time range $60480 \le {\rm MJD} \le 60580$. See \Tab~\ref{tbl:toas} for the values shown in this figure.}
  \label{fig:pulse_details}
\end{figure*}

These ToAs were then barycentred and fit for both period and DM, resulting in the values given in \Tab~\ref{tbl:ephemeris}.
The residuals are presented in \Fig~\ref{fig:pulse_details} and \Tab~\ref{tbl:toas}, along with other fitted properties of the pulses.
We also tried fitting a timing model that included a spin-down/up parameter (fixing the DM to $733\,{\rm pc}\,{\rm cm}^{-3}$), and found that it was consistent with zero.
However, owing to the recent discovery that the LPT CHIME/ILT\,J1634+44 is spinning up \citep[$\dot{P} = (-9.03 \pm 0.11) \times 10^{-12}\,{\rm s}\,{\rm s}^{-1}$;][]{2025arXiv250705139D,2025arXiv250705078B}, we report the fitted period derivative value of $(-10 \pm 92) \times 10^{-12}\,{\rm s}\,{\rm s}^{-1}$ instead of only providing an upper limit for the spin-down.

\begin{table}
  \centering
  \caption{Timing ephemeris for \src{}}
  \label{tbl:ephemeris}
  \begin{tabular}{lc}
    \hline
    Parameter & Value \\
    \hline
    Period (s) & $4186.3285 \pm 0.0002$ \\
    PEPOCH (MJD) & $59965.03792 \pm 0.00003$ \\
    DM (pc cm$^{-3}$) & $733 \pm 22$ \\
    $\tau_{\rm sc,1 GHz}$ (s) & ${\sim}0.07$ \\
    Spindown rate (${\rm s}\,{\rm s}^{-1}$) & $(-10 \pm 92) \times 10^{-12}$ \\
    \hline
  \end{tabular}
\end{table}

The variability of the pulse brightness, coupled with the fact that none of our detections were simultaneous at multiple telescopes, prohibit a direct spectral index measurement.
We note, however, that the fitted peak flux densities of the ASKAP and MeerKAT pulses are broadly consistent with the MWA flux densities at nearby epochs if an approximate spectral index of $\alpha = -2.0$ is assumed.
The bottom panel of \Fig~\ref{fig:pulse_details} shows the peak flux densities scaled to 1\,GHz under this assumption, revealing a relatively slow change of the pulses' brightness, not unlike GLEAM-X\,J1627$-$5235 \citep{2022Natur.601..526H}.
The in-band spectral index measurement of $\alpha=-3.1$ reported for the 2023 ASKAP pulse \citepalias{2024MNRAS.535..909D}, if consistent across the timespan of our observations, implies a turnover or spectral curvature at intermediate frequencies \citep[perhaps comparable with GPM\,1839$-$10;][]{2023Natur.619..487H}.

The measured DM of $733 \pm 22\,{\rm pc}\,{\rm cm}^{-3}$ is slightly higher than, but still consistent with, the in-band measurement of $710^{+200}_{-180}\,{\rm pc}\,{\rm cm}^{-3}$ given in \citetalias{2024MNRAS.535..909D}.
Because we have not attempted to simultaneously fit for both DM and scattering timescale, there will remain some small systematic error on this reported DM measurement.
Note, in particular, that the YMW16 and NE2001 models predict a slightly longer scattering time scale than the $70\,$ms that we have assumed throughout this work.
Breaking the degeneracy is possible in principle with a higher S/N average profile than what is achievable with the current data set.

For planning observations of \src{} at low frequencies ($\lesssim 300$\,MHz), care should be taken to incorporate scattering into the predicted ToAs.
The $N$th pulse since PEPOCH is predicted to arrive at the Solar System barycentre at
\begin{equation}
    {\rm MJD} = {\rm PEPOCH} + NP + \Delta t_{\rm DM}(\nu) + \Delta t_{\rm sc}(\nu),
    \label{eqn:timing}
\end{equation}
where PEPOCH and the period, $P$, are given in \Tab~\ref{tbl:ephemeris},
\begin{equation}
    \Delta t_{\rm DM}(\nu) = 4148\,{\rm s} \, \left(\frac{\rm DM}{{\rm pc\,cm}^{-3}}\right) \left(\frac{\nu}{\rm MHz}\right)^{-2}
\end{equation}
is the usual dispersion attributed to the interstellar medium, and
\begin{equation}
    \Delta t_{\rm sc}(\nu) = -\sqrt{2}\sigma\erfcx^{-1}\left(\frac{\tau(\nu)}{\sigma}\sqrt{\frac{2}{\pi}}\right) + \frac{\sigma^2}{\tau(\nu)}
    \label{eqn:scattering_delay}
\end{equation}
is the amount that the \emph{peak} of the observed pulse is delayed by scattering.
When calculating this quantity, we recommend using $\sigma = 15\,$s, a typical value for low-frequency pulses (see \Fig~\ref{fig:pulse_details}).
At frequencies above $300\,$MHz, $\tau(\nu) \ll \sigma$, and the delay due to scattering, which asymptotically behaves like $\Delta t_{\rm sc}(\nu) \sim \tau$, can be safely neglected.
At frequencies below $150\,$MHz, the asymptotic behaviour
\begin{equation}
    \Delta t_{\rm sc}(\nu) \sim \sigma \sqrt{\ln \left(\frac{\tau^2}{2\pi \sigma^2}\right)} + \frac{\sigma^2}{\tau}
\end{equation}
is accurate to within a few seconds.
At intermediate frequencies ($150\,{\rm MHz} \lesssim \nu \lesssim 300\,{\rm MHz}$), \Eqn~\ref{eqn:scattering_delay} should be used directly.



\subsection{Polarisation} \label{sec:polarisation}

No significant linear or circular polarised emission was detected in the 2023 MeerKAT pulses.
This is likely due to low S/N coupled with the source not being centred in the beam.
Much more careful imaging analysis would be needed to obtain reliable measurements or limits on the polarised components of those pulses.
We also find no significant polarised emission from the ATCA pulses.

Given \src{}'s rotation measure (RM) of ${\sim}961\,{\rm rad}/{\rm m}^2$ \citepalias[as reported in][]{2024MNRAS.535..909D}, linearly polarised emission would undergo several tens of degrees of rotation per 40\,kHz channel of the GPM observations, and therefore be somewhat depolarised.
To offer the best chance of detecting polarised emission, the MWA DDT observations between MJD 60592 and 60602 (see \Tab~\ref{tbl:obs}) were taken with a frequency resolution of 10\,kHz, which would result in only $12^\circ$ of rotation per channel at 200\,MHz.
We report no significant linearly polarised emission after de-Faraday rotating\footnote{On account of the possible sign ambiguity of the reported RM, we also tried de-Faraday rotating the MWA observations at $961\,{\rm rad}/{\rm m}^2$.} these observations at $-961\,{\rm rad}/{\rm m}^2$.
We also report that no significantly circularly polarised emission was seen in any MWA observation.

The three sufficiently bright pulses for which we report significant polarisation are (1) the original 2023 ASKAP pulse, whose polarisation has already been discussed at length in \citetalias{2024MNRAS.535..909D}, (2) the 2024 ASKAP pulse, and (3) the 2024 MeerKAT pulse.
We applied the parallactic angle correction to the MeerKAT data, as this had not already been applied during upstream processing.

\begin{figure*}
    \centering
    \includegraphics[width=0.98\linewidth]{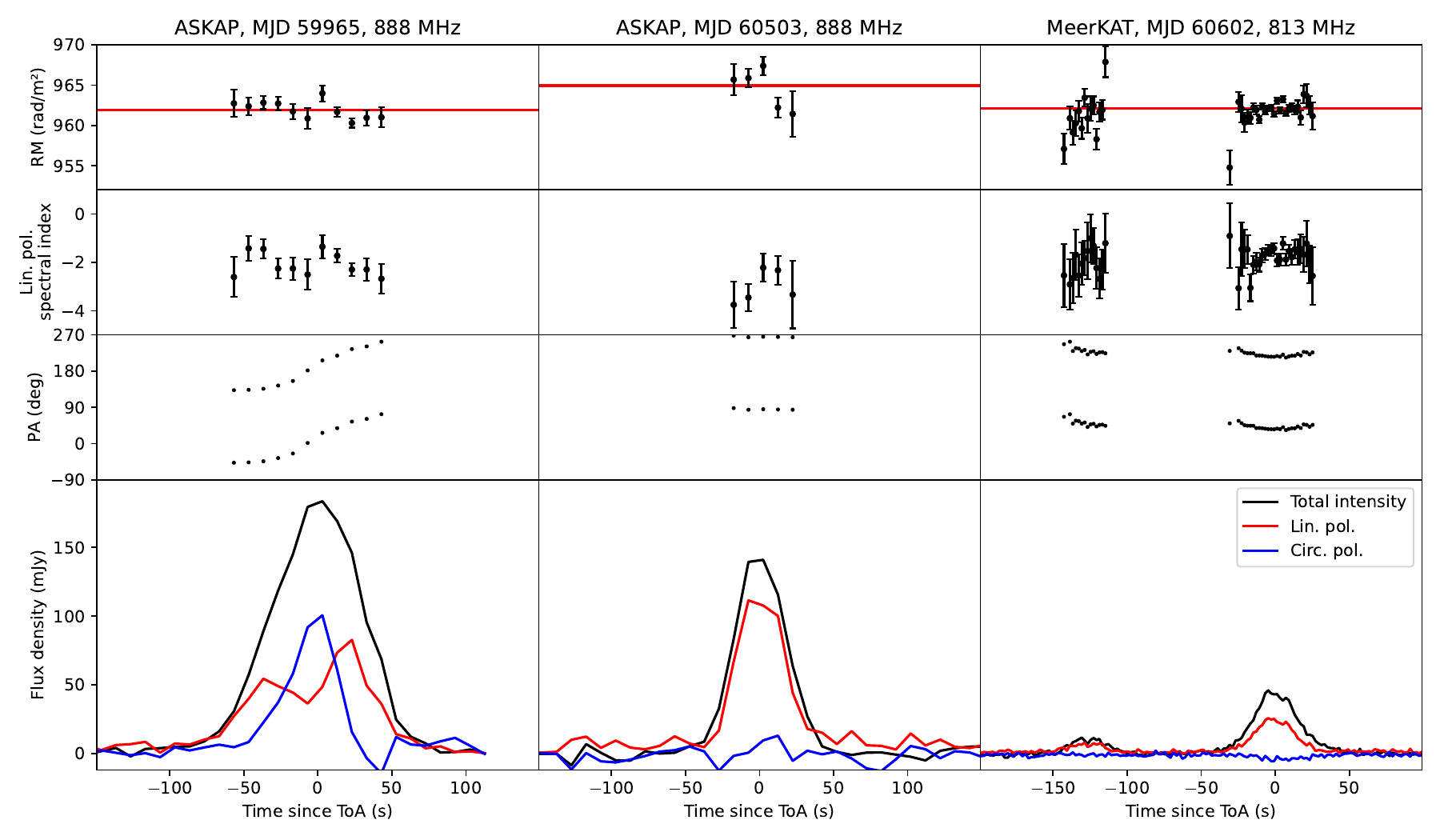}
    \caption{Polarisation of the 2023 and 2024 ASKAP pulses (left and middle) and the 2024 MeerKAT pulse (right). The top two rows show the RM and $\alpha_{\rm lin}$ parameters fitted to \Eqn~\ref{eqn:RM} for each time bin independently. Only bins for which the uncertainty on $\alpha_{\rm lin}$ is less than $1.4$ are shown, a threshold which was found to cleanly eliminate off-pulse bins. The weighted average RMs for each pulse are indicated by the horizontal red lines. The PAs and lightcurves shown in the bottom two rows of panels were produced using only the average RM for the respective pulse.}
    \label{fig:RM}
\end{figure*}

The high S/N of each time bin allowed us to test whether the RM was consistent across each pulse.
We therefore performed a joint fit of the Stokes Q and U spectra of each time bin to
\begin{equation}
    Q(\nu) = {\rm Re}(\mathcal{L}(\nu)),
    \quad
    U(\nu) = {\rm Im}(\mathcal{L}(\nu)),
    \label{eqn:RM}
\end{equation}
where
\begin{equation}
    \mathcal{L}(\nu)
        = L_{\rm 1\,GHz}\left(\frac{\nu}{\rm 1\,GHz}\right)^{\alpha_L} e^{2i({\rm RM}\,\lambda^2 + \psi_0)},
\end{equation}
$L_{\rm 1\,GHz}$ is the flux density of the linearly polarised component at 1\,GHz, $\alpha_L$ is its spectral index, RM is the rotation measure, $\lambda = c/\nu$ is the wavelength, and $\psi_0$ is the intrinsic polarisation angle.
The RMs and spectral indices of these fits are shown in the top two rows of panels of \Fig~\ref{fig:RM}.

We found that the RM derived for the MeerKAT pulse had the opposite sign, but similar magnitude, to the ASKAP pulses, which is most likely due to a different sign convention being used within the two telescopes' pre-processing software.
The agreement presented in \Fig~\ref{fig:RM} was achieved by first negating Stokes Q in the MeerKAT data.
Any reflection in the Q-U plane (e.g. ${\rm Q} \leftrightarrow -{\rm Q}$, ${\rm U} \leftrightarrow -{\rm U}$, or ${\rm Q} \leftrightarrow {\rm U}$) will achieve a similar correction to the sign of the RM, but different choices will introduce different offsets to the PA.
Because of this, the MeerKAT PA values should not be directly compared with the ASKAP values.
The most likely scenario, however, is that the MeerKAT values include an arbitrary offset of $n \times 45^\circ$, for some integer $n$.

In all cases, the magnitude of the fitted RM was within a few rad/m$^2$ of the value originally derived for the 2023 ASKAP pulse in \citetalias{2024MNRAS.535..909D}.
The RM appears to decrease slightly over the course of the 2023 ASKAP pulse, and \emph{increase} slightly over the course of the two 2024 MeerKAT components; however, all values are consistent with a constant RM across all three pulses.
The per-pulse weighted averaged values (with weighting proportional to the Stokes I profile) are $962 \pm 1$\,${\rm rad}/{\rm m}^2$ (2023 ASKAP), $965 \pm 2$ (2024 ASKAP) and $962 \pm 2$\,${\rm rad}/{\rm m}^2$ (2024 MeerKAT).
The spectral index of the linear component, $\alpha_L$, is similarly consistent, with averages value of $-2.1 \pm 0.4$ (2023 ASKAP), $-3.0 \pm 0.6$ (2024 ASKAP) and $-1.9 \pm 0.5$ (2024 MeerKAT).

Each pulse was de-Faraday rotated with the weighted average RM given above, to derive the linear polarisation profile and the PA evolution across the pulses.
For the 2023 ASKAP pulse, we recover the `S'-shaped PA curve discussed at length in \citetalias{2024MNRAS.535..909D} in the context of the rotating vector model (RVM).
Contrastingly, the PA curve of the 2024 ASKAP pulse is flat (across five 10-second bins), and that of the 2024 MeerKAT pulse is curved, but not in an RVM-like way.

\section{Discussion} \label{sec:discussion}

After accounting for dispersion and scattering effects, the timing analysis of \src{} confirms its status as an LPT, as originally conjectured in \citetalias{2024MNRAS.535..909D}.
It has the fifth largest period (${\sim}1.16\,$hours) of published LPTs to date, after ASKAP J1839$-$0756 \citep[$6.45\,$hours;][]{Lee2025}, GLEAM-X J0704$-$37 \citep[$2.92\,$hours;][]{2024ApJ...976L..21H,2025A&A...695L...8R}, ILT J1101+5521 \citep[$2.09\,$hours;][]{deRuiter2025}, and GCRT J1745$-$3009 \citep[$1.28\,$hours;][]{2005Natur.434...50H}.

With the timing solution in hand, we can explore the efficacy of the original follow-up campaign described in \citetalias{2024MNRAS.535..909D}.
Assuming \src{} remained at a similar brightness (see \Fig~\ref{fig:pulse_details}), it should have been detected in two ASKAP observations (2023-04-06 and 2023-10-18) and a few dozen times throughout the MWA's 2022 GPM run (2022-06-02 to 2022-09-08).
More careful cleaning and imaging of the 2023-04-06 ASKAP observation revealed a low-S/N pulse, at the time predicted by the ephemeris, which is presented in \Figs~\ref{fig:pulsestack} and \ref{fig:pulse_details}.
There are also three observations in the earliest part of the 2024 GPM run (one on 2024-06-02 and two on 2024-06-10) in which no pulse was observed.
However, pulses were detected just ten days later on 2024-06-20 (the first of the MWA 200\,MHz points in \Fig~\ref{fig:pulse_details}), and at each epoch thereafter whenever the MWA was observing in \src{}'s direction when a pulse was due to arrive.

In contrast, the ATCA observation taken on 2024-10-12 did not yield any detections (out of four possible predicted ToAs that occurred during that observation).
This is despite the fact that pulses were detected with ATCA just eight days prior with a similar instantaneous noise level (${\sim}1\,$mJy), as well as the fact that pulses were detected at the MWA within 24 hours of the ATCA observation (none of the MWA observations were coincident with the ATCA observation).
However, even a modest amount of pulse-to-pulse variability would allow for the possibility that the pulses during the 2024-10-12 were present, but fell below the detection threshold of ATCA.

We conclude, therefore, that \src{} is intermittent with active periods lasting on the order of months, and that the most recent activity period fortuitously started shortly after the start of the 2024 GPM observation run.
The entire set of observations presented in this work, spanning from 2023 to 2024, may represent two or three different activity windows.
This is reminiscent of the behaviour of GLEAM-X J162759.5$-$523504.3, which was `on' in January 2018, `off' in Feburary, and `on' again in March \citep{2022Natur.601..526H}.
The reappearance of \src{} suggests the possibility that other known intermittent LPTs (like GLEAM-X J162759.5$-$523504.3 and GCRT J1745$-$3009) may eventually do likewise.

As noted in \citetalias{2024MNRAS.535..909D}, optical follow-up of \src{} is hampered by its position within the Galactic plane ($b = -0.12^\circ$) and the associated extinction.
Inferring the nature of this system (white dwarf vs neutron star, isolated vs binary) from the radio pulses alone is challenging, until either the population of LPTs is better understood as a whole, or some other property of the radio signal (e.g. intermittency, polarisation) can be linked to the system properties.

\begin{figure*}
    \centering
    \includegraphics[width=0.98\linewidth]{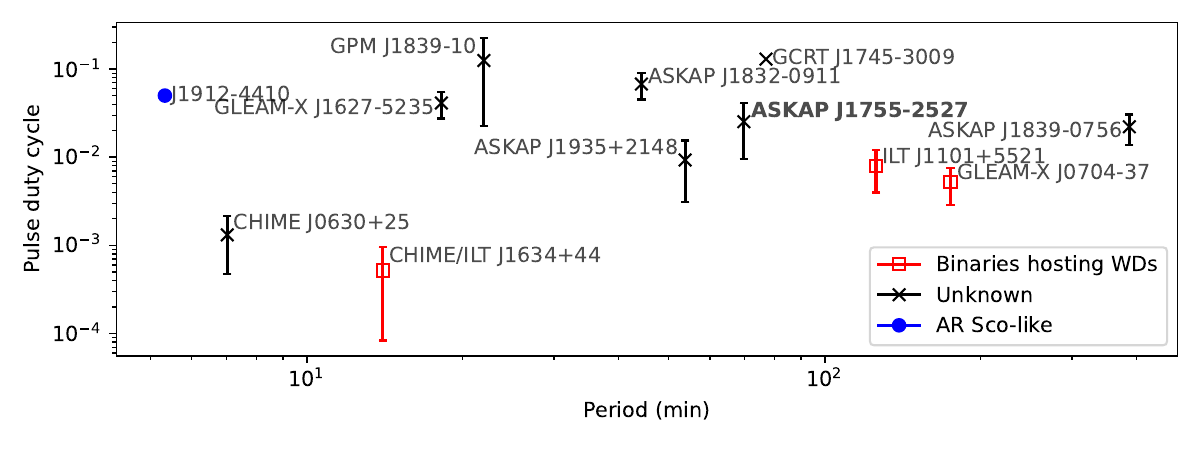}
    \caption{Duty cycle vs radio pulse period for various long period radio emitters \citep{deRuiter2025,2022Natur.601..526H,2023Natur.619..487H,2024NatAs...8.1159C,2005Natur.434...50H,deRuiter2025,Lee2025,2024arXiv241116606W}. The duty cycles are derived from the reported pulse widths for LPTs, which are compared to the range of the fitted, unscattered pulse widths of \src{} at 10\% of the peak. AR\,Sco \citep{2016Natur.537..374M,2018A&A...611A..66S} and other sources without well-defined duty cycles or periods \citep[e.g.][and references therein]{2021ApJ...920...45W} are not included on this plot.}
    \label{fig:lpt_comparison}
\end{figure*}

\src{}'s period and duty cycle are not dissimilar to ILT J1101+5521 and GLEAM-X J0704$-$37, both thought to be WD + MD binary systems (\Fig~\ref{fig:lpt_comparison}), suggesting the possibility that \src{} is also such a system.
Notably, its period of ${\sim}1.16\,$hours would make it shorter than the minimum known orbital period of $1.3\,$hours for polars \citep{2009MNRAS.397.2170G,schwope2025polarcatcatalogpolarslowaccretion}.
This makes it an interesting test bed for the possibility that LPTs with different periods represent different stages of WD + MD binary evolution.


There is as yet no clear consensus on whether LPTs represent one class or multiple classes of system \citep[e.g.][]{2024ApJ...961..214R}.
However, the recent confirmation of AR\,Scorpii \citep{2016Natur.537..374M}, J1912$-$4410 \citep{2023NatAs...7..931P}, and SDSS\,J230641.47+244055.8 \citep{2025arXiv250620455C} as white dwarf pulsars and ILT\,J1101+5521 and GLEAM-X\,J0704$-$37 as WD + MD systems \citep{deRuiter2025,2025A&A...695L...8R}, and CHIME/ILT J1634+44 as a binary hosting a white dwarf \citep[possibly two,][]{2025arXiv250705139D,2025arXiv250705078B} raise the intriguing possibility that other LPTs may also be radio-emitting white dwarf pulsars at various stages of evolution (\citealt{2021NatAs...5..648S}, Horv\'{a}th et al. submitted).

In this view, the intermittency of these systems have several possible explanations.
The limited orbital phase range in which J1912$-$4410's radio pulses were observed suggests that the geometric configuration is important in the observability of radio pulsations; `missing' pulses may simply occur at unfavourable orbital phases.
This is readily testable with long-term monitoring---this kind of intermittency will have the same periodicity as the orbit itself.

Interestingly, systems with spin-orbital (or beat-orbital) resonances of, e.g., 2:1 or 3:1, would be indistinguishable from systems in 1:1 resonances if pulses can only be observed within a limited orbital phase range.
We therefore raise the possibility that the coincidence of the observed radio periodicities of ILT\,J1101+5521 and GLEAM-X\,J0704$-$37 with their spectroscopically confirmed orbital periods does not necessarily imply that they are 1:1 systems.

In a similar vein, systems in which the spin-orbit (or beat-orbit) ratios are \emph{close to} small-integer ratios, or, similarly, systems in which precession of the WD is significant, will exhibit an extra beating effect in which an orbitally-induced intermittency is aliased to a much lower frequency.
The month-long time scale of the intermittency of \src{}, and possibly even GLEAM-X\,J1627$-$5235, may be of this kind; much longer-term monitoring is needed to confirm or refute this.

Another possibility is that radio emission is only observed when the companion produces dense enough particle outflows, in the form of winds, to activate the emission mechanism in the vicinity of the WD.
This may be the case, for example, for GLEAM-X\,J0704$-$37, whose spectroscopically detected H$\alpha$ emission is associated with the M-dwarf, and which may come from its stellar wind \citep{2025A&A...695L...8R}.
Such outflows can last for months \citep{2021ApJ...915...37W}, and would not be periodic.
We also note that the intermittency and varying polarisation properties of \src{} (and other LPTs) may be explained by changes in the local plasma environment (Rose et al., in prep).
If intermittency is of this kind, then the observed activity windows will be similarly aperiodic, or quasiperiodic.
Again, long-term monitoring can help distinguish between these scenarios.

\section{Summary} \label{sec:summary}

We have shown that \src{} is in fact an LPT, as conjectured in \citetalias{2024MNRAS.535..909D}).
Its period of ${\sim}1.16\,$hours is slightly shorter than the minimum known orbital period of optically confirmed polars, marking it as an interesting case study in the context of \citet{2025A&A...695L...8R}'s suggestion that `short'-LPTs and `long'-LPTs may correspond to different classes of CVs.

A timing solution has been presented (\Eqn~\ref{eqn:timing}) which includes a term to account for scattering at low frequencies ($\lesssim 300\,$MHz).
Without it, ToA predictions may be early by up to several tens of seconds, which is a sizeable fraction of the intrinsic pulse width.

The three higher-frequency pulses for which polarimetry was observable exhibited unique single-pulse polarisation behaviours.
Only the original ASKAP detection exhibited a PA curve that was amenable to being modelled with the RVM; the others show either flat, or slightly concave up PA curves (\Fig~\ref{fig:RM}).

We conjecture that if \src{} is a WD pulsar, its intermittency may be a result of the spin-orbital resonance being near, but not exactly, a small-integer ratio, creating a aliased beating pattern on a time scale of months.
The same explanation may be applied to other LPTs with activity cycles that last on the order of months.
An alternative explanation is that the radio emission is drived by particle wind outflows from a main sequence companion.
Long-term monitoring of LPTs will reveal whether the intermittency is periodic, and thus help discriminate between these two scenarios.

\section*{Acknowledgements}

This research is supported by an Australian Government Research Training Program (RTP) Scholarship.

N.H.-W. is the recipient of an Australian Research Council Future Fellowship (project number FT190100231).
 
This scientific work uses data obtained from Inyarrimanha Ilgari Bundara, the CSIRO Murchison Radio-astronomy Observatory. Support for the operation of the MWA is provided by the Australian Government (NCRIS), under a contract to Curtin University administered by Astronomy Australia Limited. ASVO has received funding from the Australian Commonwealth Government through the National eResearch Collaboration Tools and Resources (NeCTAR) Project, the Australian National Data Service (ANDS), and the National Collaborative Research Infrastructure Strategy.
The Australian SKA Pathfinder is part of the Australia Telescope National Facility which is managed by CSIRO (https://ror.org/05qajvd42). Operation of ASKAP is funded by the Australian Government with support from the National Collaborative Research Infrastructure Strategy. ASKAP and the MWA use the resources of the Pawsey Supercomputing Centre. Establishment of ASKAP, Inyarrimanha Ilgari Bundara, and the Pawsey Supercomputing Centre are initiatives of the Australian Government, with support from the Government of Western Australia and the Science and Industry Endowment Fund. We acknowledge the Wajarri Yamaji People as the Traditional Owners and Native Title Holders of the observatory site.

The MeerKAT telescope is operated by the South African Radio Astronomy Observatory, which is a facility of the National Research Foundation, an agency of the Department of Science and Innovation.
Observations made use of the Pulsar Timing User Supplied Equipment (PTUSE) servers at MeerKAT which were funded by the MeerTime Collaboration members ASTRON, AUT, CSIRO, ICRAR-Curtin, MPIfR, INAF, NRAO, Swinburne University of Technology, the University of Oxford, UBC and the University of Manchester.  The system design and integration was led by Swinburne University of Technology and Auckland University of Technology in collaboration with SARAO and supported by the ARC Centre of Excellence for Gravitational Wave Discovery (OzGrav) under grant CE170100004.

The database of MWA candidates, and the webapp designed to facilitate manually inspecting them, were developed by Astronomy Data and Computing Services (ADACS) as part of its software support scheme.

S.M. would like to thank J. Pritchard and E. Lenc for technical discussions regarding the ASKAP data sets included in this paper.

K.R. thanks the LSST-DA Data Science Fellowship Program, which is funded by LSST-DA, the Brinson Foundation, and the Moore Foundation; Their participation in the program has benefited this work.
\section*{Data Availability}

The raw visibilities for the MWA\footnote{\url{https://asvo.mwatelescope.org/}}, ASKAP\footnote{\url{https://research.csiro.au/casda/}}, and MeerKAT\footnote{\url{https://archive.sarao.ac.za/}} data sets are available via the respective telescopes' data portals.
The ATCA\footnote{\url{https://atoa.atnf.csiro.au/}} data is available on reasonable request to the authors, and will become publicly available from the relevant data portal by November, 2025.
The post-imaged dynamic spectra, analyses, and timing analysis data products (ToAs, pulse properties) are publicly available on GitHub\footnote{\url{https://github.com/robotopia/j1755-2527}}.
 



\bibliographystyle{mnras}
\bibliography{biblio} 




\appendix

\section{MeerKAT localization}
\label{app:localization}

\begin{figure*}
    \centering
    \includegraphics[width=0.48\linewidth]{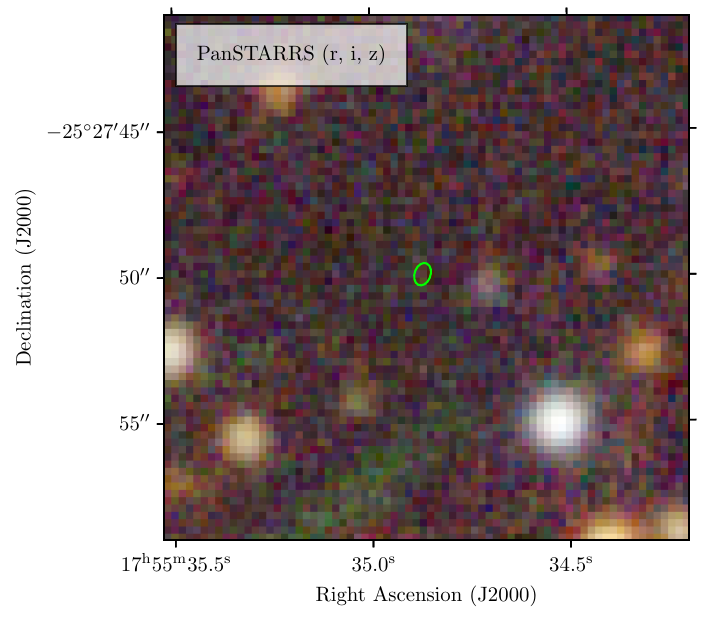}
    \includegraphics[width=0.48\linewidth]{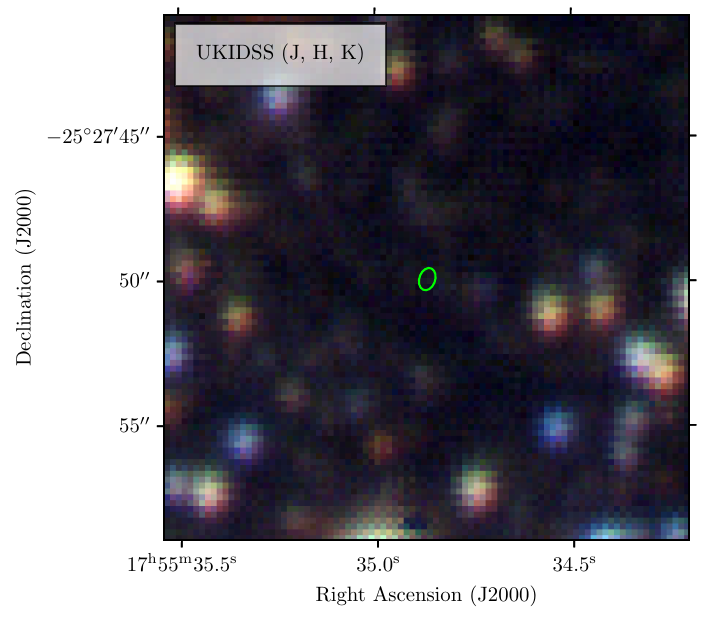}
    \caption{The $1\sigma$ positional uncertainty of \src{} derived from the MeerKAT observation taken on MJD 60602 is shown as the green ellipses overlaid on the optical (r, i, z) Panoramic Survey Telescope and Rapid Response System \citep[PanSTARRS;][]{2010SPIE.7733E..0EK,2016arXiv161205560C} image (left panel), and the infrared (J, H, K) United Kingdom Infrared Telescope Deep Sky Survey \citep[UKIDSS;][]{2007MNRAS.379.1599L} image (right panel).}
    \label{fig:localization}
\end{figure*}

\Fig~\ref{fig:localization} shows the position derived from the MeerKAT detection on MJD 60602.

\section{The effect of scattering on timing}
\label{app:scattering_dm}

It can be shown that pulses scatter-broadened by a thin screen\footnote{Relaxing the assumption of a single, thin, isotropic screen affects the shape of the scattered pulse in ways that could plausibly affect timing \citep{2009MNRAS.395.1391R}, but quantification of this is beyond the scope of this appendix.} will appear to the observer as the original pulse convolved with a one-sided exponential kernel with an associated timescale $\tau$ \citep{1972MNRAS.157...55W,1973MNRAS.163..345W}.
For intrinsically Gaussian pulses of scale $\sigma$, the result of this convolution is well described by the \textit{exponentially modified Gaussian} (EMG), as given in \Eqn~\ref{eqn:emg}.

In the regime $\tau \ll \sigma$, the effect of scattering is negligible and the scattered pulse resembles the original pulse.
ToAs can be obtained in the usual way, i.e., using a matched filter `template' constructed from the average pulse profile.

On the other hand, in the regime $\tau \gg \sigma$, the leading edge of the EMG rises sharply (resembling the error function, $\erf$) and the trailing edge resembles the exponential kernel itself.
For this reason, pulse ToAs are typically mapped to the rising edge of highly scattered pulses, which closely approximates the centre of the original unscattered pulse.

In this appendix, we discuss the effect of scattering on measuring a DM from pulse observations if scattering is not taken into account.
As will be shown, the effect is most dramatic when $\tau \approx \sigma$, and in the case of wide, highly scattered pulses from LPTs like \src{}, can lead to a discrepant DM measurement of several hundreds of pm\,cm$^{-3}$.

We consider four methods for defining ToAs:
\begin{subequations}
(1) the position where the leading edge reaches half-maximum (LEHM),
\begin{equation}
    \emg(\ToA{LEHM}) = \frac{1}{2} A\exp\left(-\frac{1}{2}\left(\frac{\mu - t_m}{\sigma}\right)^2\right),
\end{equation}
where
\begin{equation*}
    t_m = \mu - \sqrt{2}\sigma\erfcx^{-1}\left(\frac{\tau}{\sigma}\sqrt{\frac{2}{\pi}}\right) + \frac{\sigma^2}{\tau^2}
\end{equation*}
is the mode of the EMG,
(2) the inflection point on the leading edge (IPLE),
\begin{equation}
    \left.\dd{\emg(t)}{t}\right|_{t = \ToA{IPLE}} = 0,
\end{equation}
(3) the position of the peak of the pulse (PEAK),
\begin{equation}
    \ToA{PEAK} = t_m,
\end{equation}
and (4) the maximum of the convolution of the pulse with a Gaussian profile with scale $\sigma$ (TMPL)
\begin{equation}
    \left.\deriv{}{t}\left(\emg(t) \ast \exp\left[ -\frac{1}{2} \left(\frac{t - \mu}{\sigma}\right)^2 \right]\right)\right|_{t = \ToA{TMPL}} = 0.
\end{equation}
The last definition is akin to template matching with a high-frequency (negligibly scattered) pulse profile.
\end{subequations}

None of the methods is entirely accurate in the presence of scattering.
We define the errors of the ToAs to be the difference between the measured ToA and the mean of the unscattered pulse,
\begin{equation}
    \Delta \ToA{} \equiv \ToA{} - \mu.
\end{equation}
For LEHM and IPLE, the ToA is measured to be earlier than $\mu$ ($\Delta{\rm ToA} < 0$); for PEAK and TMPL, later ($\Delta{\rm ToA} > 0$).
$\ToA{LEHM}$ and $\ToA{IPLE}$ will perform much better than $\ToA{PEAK}$ and $\ToA{TMPL}$ highly scattered pulses ($\tau \gg \sigma$), whereas the converse is true when scattering is minimal ($\tau \ll \sigma$).
All methods, however, perform relatively poorly in the intermediate regime when $\tau \approx \sigma$.

\begin{figure*}
    \centering
    \includegraphics[width=0.98\linewidth]{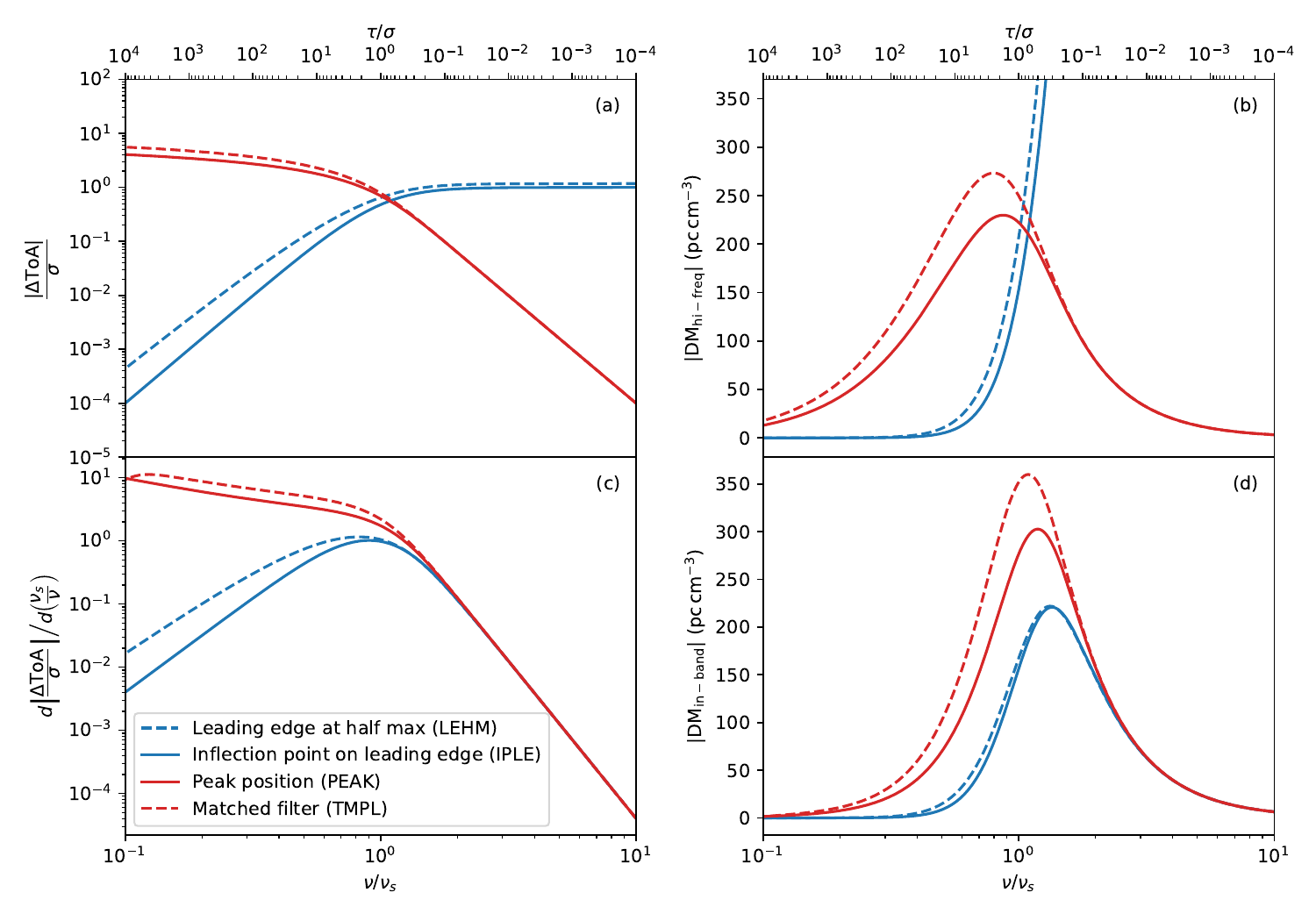}
    \caption{Analysis of DM measurement errors if scattering is not properly taken into account. Panel (a) shows the errors measured for each defined type of ToA, normalised to the scale of the original pulse, $\sigma$. In panel (b), the predicted error in DM is shown if the ToA error (compared to a high-frequency ToA measurement) is erroneously attributed to dispersion. Panel (c) shows the rate of change of the ToA error with frequency, and panel (d) shows the error in DM measured if this rate of change (the `slope' of the pulse across the observed dynamic spectrum) is erroneously attributed to dispersion. In panels (c) and (d), the derived DM errors assume a pulse scale of $\sigma = 25\,$s and $\nu_s = 230\,$MHz, as estimated for \src{}. In all panels, the blue curves indicate \emph{negative} ToA errors (i.e. the measured ToAs are earlier than the pulse's unscattered mean, $\mu$), and the red curves indicate positive quantities.}
    \label{fig:scattering_DM}
\end{figure*}

We computed numerically \citep[using SciPy's \texttt{root} function;][]{2020NatMe..17..261V} the errors for the four ToA types defined above, in the regime around $\tau \approx \sigma$.
These are shown in the panel (a) of \Fig~\ref{fig:scattering_DM}, normalised to the scale of the pulse, $\sigma$.
The axis along the top shows the normalised timescale (inverted, with largest timescales on the left), while the axis along the bottom shows the equivalent frequencies normalised to $\nu_s$, defined as the frequency at which $\tau = \sigma$.
We have assumed a scattering index of $-4$, such that
\begin{equation}
    \frac{\tau}{\sigma} = \left(\frac{\nu}{\nu_s}\right)^{-4}.
\end{equation}

A set of ToAs may be converted to DM measurements in two ways: (1) by comparing the ToAs to those measured at much higher frequencies (e.g. with TMPL) where the scattering is known to be negligible, and (2) by measuring the slope of the pulse in the dynamic spectrum (for example, by measuring the ToAs in subbands across the observed frequency range).
Panel (c) of \Fig~\ref{fig:scattering_DM} shows the expected slope measurements for the ToA errors given in the panel (a).

In the first case, the ToA error of a pulse measured at frequency $\nu$, if attributed erroneously to the effect of dispersion, will yield a discrepant DM of
\begin{equation}
    \Delta{\rm DM}_{\rm hi-freq} \approx \frac{\Delta\ToA{}\,\nu^2}{\mathcal{D}}.
\end{equation}
For the specific case of \src{}, for which we estimate $\sigma = 25\,$s and $\nu_s = 230\,$MHz, consistent with $\tau_{\rm sc,1 GHz} = 70\,$ms, the DM error can be as high as a few hundreds of pc\,cm$^{-3}$.
When using this DM method, it is better to use LEHM or IPLE at frequencies $\nu \lesssim \nu_s$, and PEAK or TMPL for $\nu \gtrsim \nu_s$; however, it should be remembered that LRHM and IPLE will \emph{underestimate} the DM, while PEAK and TMPL will \emph{overestimate} it.
All ToA definitions produce sizeable DM errors in the approximate range $\nu_s \lesssim \nu \lesssim 3\nu_s$.

The situation is not much improved for DMs determined via the slope of the pulse across the observed band, for which the scattering-induced slope is mapped to an (erroneous) equivalent DM measurement of
\begin{equation}
    \Delta{\rm DM}_{\rm in-band} \approx \deriv{{\rm ToA}}{\nu} \frac{\nu^3}{2\mathcal{D}}
\end{equation}
The calculated numbers for \src{} are illustrated in panel (d).
For this case, LEHM and IPLE are always preferred to PEAK and TMPL, but the magnitude of the derived DM error is still in the hundreds of pc\,cm$^{-3}$ at frequencies in the range $\nu_s \lesssim \nu \lesssim 3\nu_s$.

We confirm the order of magnitude of the above results by reporting a measured in-band DM for one of the MWA pulses at 185\,MHz, using the TMPL method to measure ToAs in 1.28\,MHz subbands from ${\sim}170$ to ${\sim}200\,$MHz, of ${\rm DM}_{\rm in-band} = 1221\pm257\,$pc\,cm$^{-3}$.
This exceeds the DM reported in this work by ${\sim}450\,$pc\,cm$^{-3}$, in excess of the prediction shown in panel (d), but consistent within measurement errors.

\section{Pulse properties}

\Tab~\ref{tbl:toas} provides the fitted parameter values of the pulses shown in \Fig~\ref{fig:pulse_details}.

\begin{table*}
  \centering
  \caption{Times of arrival and other fitted pulse properties}
  \label{tbl:toas}
  \begin{tabular}{cccccc}
\hline
Telescope & Freq & ToA & $S_{\rm peak}$ & $\sigma$ & Fluence \\
 & (MHz) & (MJD) & (Jy) & (s) & (Jy\,s) \\
\hline
ASKAP & 888 & 59965.0429346(42) & 0.184(5) & 29.1(4) & 13.4(2) \\
MWA & 170 & 59966.061584(40) & 4(1) & 40(7) & 358(55) \\
ASKAP & 888 & 60040.913447(32) & 0.015(3) & 25(3) & 1.0(1) \\
MeerKAT & 1284 & 60092.8995476(100) & 0.0032(4) & 12.8(9) & 0.104(7) \\
MeerKAT & 1284 & 60092.947966(10) & 0.0033(6) & 12(1) & 0.10(1) \\
MeerKAT & 1284 & 60092.996466(11) & 0.0032(3) & 15.8(9) & 0.125(7) \\
MeerKAT & 1284 & 60093.044919(12) & 0.0030(3) & 16(1) & 0.122(7) \\
MWA & 200 & 60481.637466(65) & 0.8(4) & 24(8) & 50(11) \\
MWA & 200 & 60485.658977(33) & 0.7(2) & 16(4) & 29(3) \\
MWA & 200 & 60489.632266(20) & 1.4(3) & 14(2) & 48(3) \\
MWA & 200 & 60492.588020(24) & 2.4(6) & 14(3) & 87(7) \\
MWA & 200 & 60496.609639(18) & 1.5(3) & 17(2) & 67(4) \\
MWA & 200 & 60503.587085(14) & 2.0(2) & 17(2) & 82(3) \\
ASKAP & 888 & 60503.6347610(51) & 0.151(7) & 16.4(5) & 6.2(2) \\
MWA & 200 & 60506.542809(33) & 2(1) & 9(4) & 57(7) \\
MWA & 200 & 60510.564531(11) & 2.0(2) & 16(1) & 79(3) \\
MWA & 200 & 60519.528813(29) & 2.4(8) & 14(4) & 86(10) \\
MWA & 200 & 60522.533028(16) & 1.6(2) & 18(2) & 76(3) \\
MWA & 200 & 60527.524033(14) & 1.8(2) & 16(2) & 73(3) \\
MWA & 200 & 60530.528339(16) & 1.6(2) & 17(2) & 68(4) \\
MWA & 200 & 60537.506042(13) & 2.1(3) & 17(2) & 92(4) \\
MWA & 200 & 60541.479447(47) & 1.1(5) & 14(5) & 41(6) \\
MWA & 200 & 60544.483761(15) & 1.6(2) & 20(2) & 79(3) \\
MWA & 200 & 60551.461625(16) & 2.2(3) & 16(2) & 85(4) \\
MWA & 185 & 60572.395455(15) & 1.9(3) & 15(2) & 74(3) \\
MWA & 185 & 60572.443943(21) & 1.8(3) & 16(2) & 73(4) \\
MWA & 185 & 60572.540824(18) & 1.6(3) & 16(2) & 64(4) \\
MWA & 185 & 60572.589281(31) & 1.6(4) & 21(4) & 82(8) \\
ATCA & 2100 & 60587.270973(26) & 0.003(1) & 12(2) & 0.10(2) \\
ATCA & 2100 & 60587.319527(28) & 0.003(1) & 10(3) & 0.09(2) \\
MWA & 200 & 60592.456765(27) & 0.7(2) & 19(3) & 34(3) \\
MWA & 200 & 60592.505159(33) & 0.7(2) & 19(4) & 35(4) \\
MWA & 200 & 60593.474369(26) & 0.8(2) & 15(3) & 29(3) \\
MWA & 200 & 60593.522901(46) & 0.6(2) & 19(5) & 27(4) \\
MWA & 200 & 60594.443484(21) & 0.9(2) & 18(2) & 43(3) \\
MWA & 200 & 60594.491990(33) & 0.6(3) & 12(4) & 19(2) \\
MWA & 200 & 60595.461118(29) & 0.7(2) & 19(3) & 33(3) \\
MWA & 200 & 60595.509617(25) & 0.8(2) & 13(3) & 27(2) \\
MWA & 200 & 60596.430283(20) & 1.0(2) & 16(2) & 39(3) \\
MWA & 200 & 60596.478721(23) & 0.9(2) & 17(3) & 39(3) \\
MWA & 200 & 60597.447846(41) & 0.5(2) & 16(5) & 22(3) \\
MWA & 200 & 60597.496319(30) & 0.8(2) & 19(3) & 38(4) \\
MWA & 200 & 60599.434635(17) & 1.2(2) & 13(2) & 39(2) \\
MWA & 200 & 60599.483122(24) & 1.0(3) & 13(3) & 32(3) \\
MWA & 200 & 60600.500655(31) & 0.7(2) & 16(4) & 30(3) \\
MWA & 200 & 60602.438964(32) & 0.7(2) & 21(4) & 36(4) \\
MWA & 200 & 60602.487394(27) & 0.8(2) & 20(3) & 39(3) \\
MeerKAT & 813 & 60602.5835733(40) & 0.044(2) & 14.9(4) & 1.64(4) \\
\hline
  \end{tabular}
\end{table*}

\bsp	
\label{lastpage}
\end{document}